
\magnification=1200
\baselineskip=12pt
\input tables
\def\lsim{<\kern-2.5ex\lower0.85ex\hbox{$\sim$}\ }
\def\rsim{>\kern-2.5ex\lower0.85ex\hbox{$\sim$}\ }
\overfullrule=0pt
\def\LAMBDABAR {\hbox{$\lambda$\kern-0.52em\raise+0.45ex\hbox{--}\kern+0.2em}}
\def\ebar {\hbox{E\kern-0.6em\raise0.2ex\hbox{/}\kern+0.1em}}
\rightline{UR--1235$\ \ \ \ \ \ $}
\rightline{ER40685--687}
\baselineskip=20pt
\centerline{\bf NONLEPTONIC DECAYS OF CHARMED MESONS}
\centerline{\bf INTO TWO PSEUDOSCALAR MESONS}
\vskip 2cm
\centerline{Ashok Das}
\centerline{and}
\centerline{Vishnu S. Mathur}
\centerline{Department of Physics and Astronomy}
\centerline{University of Rochester}
\centerline{Rochester, NY 14627}
\vskip 2cm
\centerline{\underbar{Abstract}}

We investigate the nonleptonic decay of charmed meson into two pseudoscalar
mesons using the vector--dominance model, and compare the results with
those obtained from the factorization model.  In particular, we discuss the
role of the annihilation diagrams in the two models.
\vfill\eject

Much of what we understand
 about nonleptonic decays of strange particles comes
from current algebra and the soft pion technique.  Unfortunately, this
technique cannot be applied to the decay of heavy mesons carrying the $c$
or the $b$ quark, since the emitted pions are generally not soft.  In this
case a completely different method has been developed based on the
assumption that the hadronic matrix element of currents factorizes$^1$.

Historically,
 the nonleptonic decays of strange particles have also been
discussed in terms of a dymanical model$^2$ based on the idea of vector
dominance.  In particular, it is well--known that for the $K \rightarrow
2 \pi$ decays this model provides a satisfactory description.  It would be
of interest to see how vector dominance fares in describing also the heavy
meson decays.  In this paper we deal with this question and, as an
extension of $K \rightarrow 2 \pi$, discuss the decay of $D$ and $D_S$
mesons into two pseudoscalars.  Indeed such an analysis was attempted
several years ago$^3$.  The present work obviously benefits from the
 availability of better data.  More importantly, however, we undertake a
detailed comparison of the vector dominance model and the factorization
model.  In particular, we discuss the role of the so--called annihilation
amplitudes in the two models.

For nonleptonic decay of charm, the effective weak Hamiltonian in the
current--current form may be taken to be$^4$
$$H_W = {G_F \over \sqrt{2}} \left[ a_1 (\overline u d^\prime)_\mu
(\overline s^\prime c )_\mu + a_2 (\overline s^\prime d^\prime)_\mu
 (\overline u c)_\mu \right] \eqno(1)$$
where $(\overline q_\beta q^\alpha)_\mu$ are color--singlet V--A currents
$$(\overline q_\beta q^\alpha )_\mu = i
\overline q_\beta \gamma_\mu (1 + \gamma_5 ) q^\alpha =
 (V_\mu)^\alpha_\beta + (A_\mu)^\alpha_\beta
\qquad (\alpha, \beta = 1,2 \dots 4) \eqno(2)$$
and $a_1,\ a_2$ are real coefficients which we treat as phenomenological
parameters.  The primed quark fields are related to the unprimed ones by
the usual Cabibbo--Kobayashi--Maskawa (CKM) mixing matrix.  We shall ignore
the Penguin--type contributions.  For the nonleptonic $D$ and $D_S$ decays
into two mesons, the Hamiltonian (1) leads to two main classes of
quark--model diagrams, the spectator and the annihilation diagrams shown in
Figs. 1(a) and 1(b) respectively.  It is well--known that the
 annihilation--diagram contribution in the quark model is helicity
suppressed.

The factorization model assumes that the matrix element for the decay $D
 \rightarrow P_1 P_2 \ (P_1$ and $P_2$ are light mesons) can be written
in the factorized form
$$\eqalign{< P_1 (q_1) &P_2 (q_2) \vert H_W \vert D(q)>\cr
 \sim &{G_F \over \
\sqrt{2}} \big[ a_1 <P_2 (q_2) \vert
(\overline u d^\prime)_\mu \vert 0><P_1 (q_1)
\vert (\overline s^\prime c)_\mu
\vert D(q)>\cr
&+ a_2 <P_1 (q_1) \vert (\overline s^\prime d^\prime )_\mu \vert 0>
<P_2 (q_2) \vert (\overline uc)_\mu \vert D(q)> \big]\cr}\eqno(3)$$
The two terms on the right can easily be seen to correspond to the
quark--model spectator diagrams of Fig. 1(a).  An important feature of the
factorization approximation is that the annihilation diagram can be
neglected$^1$.  To see this, note that the annihilation diagram corresponds to
the factorization
$${G_F \over \sqrt{2}} \ a_2 <0 \vert (\overline uc)_\mu
\vert D(q)><P_1 (q_1) P_2 (q_2) \vert
(\overline s^\prime d^\prime )_\mu \vert 0> \eqno(4)$$
 if the decaying particle is a neutral $D^o$.  For the charge--carrying
$D^+$ or $D^+_S$ a similar factorization can be written down involving the
$a_1$ term in the Hamiltonian.  In these factorized forms, the charmed
meson is connected to vacuum by an appropriate current.  From Lorentz
invariance, this matrix element is proportional to
$q_\mu = (q_1 + q_2)_\mu$.  Multiplied by the other matrix element in (4),
we see that the annihilation contribution is then proportional to the
matrix element of the divergence of the current $(\overline s^\prime
d^\prime)_\mu$ formed from light quarks, and is thus proportional to the
masses of the light quarks.  This is the analogue of helicity suppression
in the quark--model annihilation diagram.  By comparison, the spectator
contribution (3) is proportional to quark masses involving the heavy charmed
quark.

There is another argument that shows an additional suppression of the
annihilation contribution.  To appreciate this, note that the structure of
the matrix element $<P_1 (q_1)P_2 (q_2) \vert (\overline s^\prime d^\prime
)_\mu \vert 0>$ in (4) can be written from Lorentz invariance in terms of
form--factors to be evaluated at the momentum transfer
$(q_1 + q_2)^2 = q^2 = - m^2_D$.  In the standard pole--dominated form, the
form--factor is then expected to lead to a further suppression of the
annihilation amplitude.  By contrast, again, the form--factors appearing
in the spectator amplitude (3) are to be evaluated at the low mementum
transfer $(q - q_1)^2 = q^2_2 = - m^2_2 \ {\rm or}\ (q - q_2)^2 =
 q^2_1 = - m^2_1$, and thus need not be suppressed.

Consider now the vector--dominance model.  Here we take the currents in $H_
W$ to be the hadronic currents given by the field--current identities
 $(\alpha, \beta = 1,2 \dots 4)$
$$\eqalign{(V_\mu)^\alpha_\beta &= \sqrt{2}\ g_V (\phi_\mu)^\alpha_\beta\cr
(A_\mu)^\alpha_\beta &= \sqrt{2} \  f_P \partial_\mu P^\alpha_\beta\cr}
\eqno(5)$$
where $(\phi_\mu)^\alpha_\beta$ and $P^\alpha_\beta$ are the field
operators of the vector and the pseudoscalar mesons respectively, and
$g_V$ and $f_P$ are the corresponding decay constants.  The nonleptonic
weak interaction (1) can then be represented by a two--meson vertex.  For
the parity violating decay $D \rightarrow P_1 P_2$, the Cabibbo allowed and
once--suppressed weak vertices are listed in Figs. 2(a) and 2(b)
respectively.

The Feynman diagrams for the decay $D \rightarrow P_1P_2$ are depicted in
Fig. 3.  It is easily seen that the diagrams in Fig. 3(a) and 3(b) are
respectively the analogues of the spectator and the annihilation diagrams
of the quark model.  The annihilation amplitude in Fig. 3(b) is
proportional to
$$A_a \propto q_\mu \ {\delta_{\mu \nu} + q_\mu q_\nu /m^2_V \over
q^2 + m^2_V} \ (q_1 - q_2)_\nu \eqno(6)$$
In the pole term, if we set $q^2 = - m^2_D$, at first sight it seems to
lead to a suppression in the annihilation amplitude.  However it is trivial
to see that the pole actually cancels, and one obtains
$$A_a \propto {m^2_2 - m^2_1 \over m^2_V} \eqno(7)$$
The annihilation amplitude does depend on the masses $m_1, \ m_2$ of the
light mesons.  Again, this is the analogue of the helicity suppression in
the quark model.  However, it is easy to see that $V$ is a light vector
meson in this case, so the annihilation amplitude is actually proportional
to the ratio of light meson mass squares.  By contrast, for the spectator
diagram of Fig. 3(a), the amplitude is proportional to
$$A_s \propto (q + q_1)_\mu \ {\delta_{\mu \nu} + q_{2 \mu} q_{2 \nu}/
m^2_V \over q^2_2 + m^2_V} \ q_{2 \nu} = {m^2_1 - m^2_D \over
m^2_V} \eqno(8)$$
This time since $V$ is a charmed vector meson $D^*$ or $D^*_S$, we find
that the spectator contribution while containing a piece proportional to
the charmed meson mass term $m^2_D$, is actually determined by the ratio of
heavy meson mass squares.  Thus, there is no apriori reason why $A_a$ and
$A_s$ may not be of the same order of magnitude, and we are not justified
in neglecting the annihilation amplitude.

We now use the vector dominance model to compute the decays
$D,\ D_S \rightarrow P_1P_2$, taking into account contributions from both
the Feynman diagrams in Fig. 3.  We follow Bauer et al.$^1$ (BSW) and first
determine the parameters $a_1,\ a_2$ by confronting our model to the data
on
$D \rightarrow K \pi$ decays.  As in BSW, we have to take the final state
interaction into account, and we do this by considering only elastic
scattering in the final $K \pi$ state.  In terms of the isospin amplitudes,
we have
$$\eqalign{A(D^o \rightarrow K^- \pi^+) &= {1 \over \sqrt{3}} \ A_{3/2}
+ \sqrt{{2 \over 3}} \ A_{1/2} \cr
A(D^o \rightarrow \overline K^o \pi^o) &=
 \sqrt{{2 \over 3}}\ A_{3/2} - {1 \over \sqrt{3}} \ A_{1/2}\cr
A(D^+ \rightarrow \overline K^o \pi^+) &= \sqrt{3}\ A_{3/2}\cr}\eqno(9)$$
where
$$A_I = \vert A_I \vert e^{i \delta_I}\eqno(10)$$
is the amplitude in the isospin state $I$ and $\delta_I$ is the phase shift
in that channel.  Using the data from the recent particle properties data$^
5$ booklet, we obtain
$$\eqalign{\vert A_{1/2} \vert &= 2.94 \times 10^{-6} \ {\rm GeV}\cr
\vert A_{3/2} \vert &= 7.37 \times 10^{-7} \ {\rm GeV}\cr
\delta_{1/2} - \delta_{3/2} &= 93.4^\circ\cr}\eqno(11)$$

We now use the vector dominance model to compute these amplitudes.  Since
we do not know all the strong interaction coupling constants, we use flavor
SU(4) symmetry to relate these to $g_{\rho \pi \pi}$, which is known from
the $\rho \rightarrow 2 \pi$ decay ($g_{\rho \pi \pi} \simeq 4.0$).  The
 $D \rightarrow K \pi$ amplitudes can then be written down from the Feynman
diagrams of Fig. 3 to be
$$\eqalign{A(D^o \rightarrow K^- \pi^+) &=
\gamma \bigg[ -a_2 g_{K^*} f_D \ {m^2_\pi - m^2_K \over
m^2_{K^*}} + a_1 g_{D^*_S} f_\pi \
{m^2_K - m^2_D \over m^2_{D^*_S}} \bigg]\cr
A(D^o \rightarrow \overline K^o \pi^o) &=
{\gamma \over \sqrt{2}}\  \bigg[ a_2 g_{K^*} f_D \ {m^2_\pi - m^2_K \over
m^2_{K^*}} + a_2 g_{D^*} f_K \
{m^2_\pi - m^2_D \over m^2_{D^*}} \bigg]\cr
A(D^+ \rightarrow K^o \pi^+) &=
\gamma \bigg[ a_2 g_{D^*} f_K \ {m^2_\pi - m^2_D \over
m^2_{D^*}} + a_1 g_{D^*_S} f_\pi \
{m^2_K - m^2_D \over m^2_{D^*_S}} \bigg]\cr}\eqno(12)$$
where
$$\gamma = i \ {G_F \over \sqrt{2}}\ V^*_{cs} V_{ud} \ 2g_{\rho \pi \pi}
\eqno(13)$$
The $V$'s are the CKM matrix elements and the couplings $g_V$ and $f_P$ in
(12) are defined in (5).  For the vector meson decay constant $g_V$, we use
the spectral function sum--rule$^6$ based on asymptotic SU(4), which
predicts identical values of $g_V/m_V$ for $V = \rho,\ K^*,\ D^*,\
{\rm and}\ D^*_S$.  For the $\rho$--meson, we extract from $\rho
\rightarrow \ell \overline \ell$ decay,
 $g_\rho / m_\rho \simeq 0.15$ GeV.  For the pseudoscalar meson decay
constant $f_P$, we take $f_\pi = 0.093$ GeV, $f_K \simeq 1.2 \ f_\pi =
0.112$ GeV and choose $f_D \simeq f_{D_S} \simeq 0.136$ GeV as determined$^
7$ from the sum--rule technique in quantum chromodynamics.  The isospin
amplitudes in the vector dominance model can then be computed to be
$$\eqalign{\vert A_{1/2} \vert &= 1.07 \vert
(a_1 - 1.03 a_2) \vert \times 10^{-6}\ {\rm GeV}\cr
\vert A_{3/2} \vert &= 7.58 \vert (a_1 +
 1.34 a_2) \vert \times 10^{-7} \ {\rm GeV}\cr}\eqno(14)$$
Using the values (11) for these amplitudes obtained from the data, we find
two solutions (only the relative sign of $a_1$ and $a_2$ is important)
\medskip
\item{I.} $a_1 = 1.98 \quad , \quad a_2 = -0.75$ \hfill (15)
\medskip
\item{II.} $a_1 = 1.13 \quad , \quad a_2 = - 1.57$ \hfill  (16)
\medskip
\noindent It should be emphasized that the annihilation contribution has
been included in this analysis.  We find however that in the $ D
\rightarrow K \pi$ decays, this contribution is numerically small compared
with the spectator contribution.

Isospin analyses similar to the one above can also be performed for the
Cabibbo suppressed decay modes $D \rightarrow K \overline K$ and $D
\rightarrow \pi \pi$.  In the $D \rightarrow \pi \pi$ decays, only $D^o
\rightarrow \pi^+ \pi^-$ has a measured branching ratio at present.  It is
not hard to see that this experimental value is already sufficient to rule
out the solution II.  Also the surviving solution I is consistent with the
data with negligible final state interaction.  In the $D \rightarrow
 K \overline K$ decays, on the other hand, we find that the data require
sizable final state interaction.  In this case, while neither of the
solutions can be ruled out, we find that I satisfies the data better.  We
do not present the details of these analyses here, but accept
 the conclusion that the
present data favors the solution I over II.

It is of interest to compare our solution to the one obtained by BSW in the
factorization approach.  Using the values (11) of the isospin amplitudes
obtained from the recent $D \rightarrow K \pi$ data, we find that the BSW
solution$^8$ is given by
$$a_1 = 1.09 \quad , \quad a_2 = -0.48 \eqno(17)$$
Our individual values of $a_1$ and $a_2$ in solution I are somewhat larger,
although the ratio is not very different from the one in (17).

We have calculated the various allowed and suppressed decays of the type
$D,\ D_S \rightarrow P_1 P_2$, and have listed the results of the vector
dominance model in Table 1.  Also listed in the table are the results from
the factorization model, and the experimental data, where available.

In conclusion, we would like to emphasize that the vector dominance model
affords a natural way of taking into account the annihilation contribution.
 While this contribution is small for the $D,\ D_S \rightarrow PP$
 decays, one would expect it to be larger in decays involving
 the heavier vector mesons as in
$D,\ D_S \rightarrow PV$ and
 $D,\ D_S \rightarrow VV$.  In these cases, the vector dominance model has
to be extended to include pseudoscalar and other meson poles.  This work
will be reported elsewhere.

This work was supported in part by the U.S. Department of Energy Grant No.
 DE--FG--02--91ER40685.
\vfill\eject
\noindent {\bf \underbar{References and Footnotes}}
\item{1.} M. Bauer, B. Stech and M. Wirbel, Z. Phys. C -- Particles and
Fields {\bf 34}, 103 (1987).
\item{2.} J. J. Sakurai, Phys. Rev. {\bf 156}, 1508 (1967).
\item{  } G. S. Guralnik, V. S. Mathur and L. K. Pandit, Phys. Rev.
 {\bf 168}, 1866 (1968).
\item{3.} V. S. Mathur, Proceedings of a Conference on Quark Confinement
and Field Theory, Rochester, NY 1976, edited by D. R. Stump and D. H.
Weingarten (published by John Wiley \& Sons).
\item{4.} K. Jagannathan and V. S. Mathur, Nucl. Phys. {\bf B171}, 78
(1980) and ref. 1.
\item{5.} Review of Particle Properties, Phys. Rev. {\bf D45}, Part 2 (June
1992).  We have ignored the errors quoted in the data for our analysis
here.
\item{6.} S. Weinberg, Phys. Rev. Lett. {\bf 18}, 507 (1967).
\item{  } T. Das, V. S. Mathur and S. Okubo, Phys. Rev.
Lett. {\bf 18}, 761 (1967) and ref. 3.
\item{7.} V. S. Mathur and M. T. Yamawaki, Phys. Rev. {\bf D29}, 2057
(1984).
\item{8.} BSW have used slightly different values of the couplings $f_P$
and $g_V$ than those used by us.
\vfill\eject
\midinsert
\thicksize=.5pt
\thinsize=.5pt
\tablewidth=6in
\begintable
Branching | Vector--dominance | Factorization | Experiment$^5$ \crnorule
Ratio | Model | Model$^1$ |        \cr
BR($D^o \rightarrow K^- \pi^+$) | 4.9\% | 5.0\% | (3.65 $\pm$ 0.21)\%\cr
BR($D^o \rightarrow \overline K^o \pi^o$) | 0.8\% |
0.7\% | (2.1 $\pm$ 0.5)\%\cr
BR($D^+ \rightarrow \overline K^o \pi^+$) | 2.7\% |
2.7\% | (2.6 $\pm$ 0.4)\%\cr
BR($D^+_S \rightarrow \overline K^o K^+$) | 1.2\% | 1.3\% |
(2.8 $\pm$ 0.7)\%\cr
BR($D^+_S \rightarrow \eta \pi^+$) | 3.7 \% | 2.6\% |
(1.5 $\pm$ 0.4)\%\cr
BR($D^o \rightarrow \overline K^o \eta$) | 0.1 \% | 0.3\% |
$<$2.3\%\cr
 | | | \cr
BR($D^o \rightarrow K^+ K^-$) | 2.9 $\times$ 10$^{-3}$  | 3.7 $\times
10^{-3}$ |
(4.1 $\pm$ 0.4) $\times 10^{-3}$\cr
BR($D^o \rightarrow K^o \overline K^o$)  | 0 | 0 |
(1.1 $\pm$ 0.4) $\times 10^{-3}$\cr
BR($D^+ \rightarrow K^+ \overline K^o$) | 7.4 $\times$ 10$^{-3}$  |
 9.6 $\times
10^{-3}$ |
(7.3 $\pm$ 1.8) $\times 10^{-3}$\cr
BR($D^o \rightarrow \pi^+ \pi^-$) | 2.9 $\times$ 10$^{-3}$  | 2.6 $\times
10^{-3}$ |
(1.63 $\pm$ 0.19) $\times 10^{-3}$\cr
BR($D^o \rightarrow \pi^o \pi^o$) | 0.8 $\times$ 10$^{-4}$  | 2.5 $\times
10^{-4}$ |
$<$4.6 $\times 10^{-3}$\cr
BR($D^+ \rightarrow \pi^+ \pi^o$) | 1.4 $\times$ 10$^{-3}$  | 1.1 $\times
10^{-3}$ |
$<$5.3 $\times 10^{-3}$\cr
BR($D^+_S \rightarrow K^o \pi^+$) | 4.6 $\times$ 10$^{-3}$  | 2.3 $\times
10^{-3}$ |
$<$ 6 $\times 10^{-3}$\cr
BR($D^+_S \rightarrow K^+ \pi^o$) | 5.4 $\times$ 10$^{-4}$  | 2.3 $\times
10^{-4}$ |
\endtable
\vskip .1in
\endinsert\noindent
\centerline{\bf Table 1}
\vfill\eject
\noindent {\bf \underbar{Figure and Table Captions}}
\item{Fig. 1} Quark--model diagrams for the nonleptonic decay of $D$ and
$D_S$ into two mesons.
\item{      } 1(a) describe the spectator diagrams and 1(b) the
annihilation diagrams.
\item{Fig. 2} Parity--violating weak vertices in vector--dominance model.
2(a) describe the Cabibbo--allowed vertices and 2(b) the once--suppressed
ones.
\item{Fig. 3} Feynman diagrams for the decay $D,\ D_S \rightarrow
 P_1P_2$ in vector--dominance model.  The weak vertex is represented by a
black dot and the strong vertex by an open circle.
\item{Table 1} Branching ratios of the Cabibbo--allowed and
once--suppressed decays of the type $D,\ D_S \rightarrow P_1P_2$.  The
calculated values in the table ignore the final state interactions.  For
the vector--dominance model, we have used the values of $a_1$ and $a_2$
given in solution I.  For the factorization model, we have used the BSW
values given in (17).
\end